\documentclass{ws-ijmpd}

\def\lsim{\lower.5ex\hbox{$\; \buildrel < \over \sim \;$}}
\def\gsim{\lower.5ex\hbox{$\; \buildrel > \over \sim \;$}}
\def \simeq{\lower.3ex\hbox{$\; \buildrel \sim \over - \;$}}
\def\ch{\lower-0.55ex\hbox{--}\kern-0.55em{\lower0.15ex\hbox{$h$}}}
\def\lh{\lower-0.55ex\hbox{--}\kern-0.55em{\lower0.15ex\hbox{$\lambda$}}}
\def\eg{{\it e.g.,} }

\def\ie{{\em i.e.,} }
\def\ee{{$e^--e^+$}}
\def\ep{{$e^--p^+$}}

\begin{document}

\markboth{Indranil Chattopadhyay and Sandip K. Chakrabarti}
{Effects of flow composition on accretion properties}

%
\catchline{}{}{}{}{}
%

\title{Effects of the composition on transonic properties of accretion flows around black holes}

\author{Indranil Chattopadhyay$^1$ and Sandip K. Chakrabarti$^{2,3}$}

\address{1. ARIES, Manora Peak, Nainital-263129, Uttarakhand, India; indra@aries.res.in\\
2. S.N. Bose National Centre for Basic Sciences, JD Block, Kolkata 700098; chakraba@bose.res.in\\
3. Indian Centre for Space Physics, Chalantika 43, Garia Station Rd. Kolkata 700084 }

\maketitle

\begin{history}
\received{Day Month Year}
\revised{Day Month Year}
\end{history}

\begin{abstract}
We study the properties of a steady, multi-species, low angular momentum 
accretion flow around a Schwarzschild black hole. Each species is 
described by a relativistic equation of state. We find that the transonic 
properties depend strongly on the composition of the flow.  We find that an electron-positron
pair plasma is the least relativistic one. This flow produces only one sonic
point very close to the event horizon and does not show multiple critical points
for any angular momentum or energy. When the baryons are present, the number of
critical points depend on the specific energy content. Since the number of critical points
decide whether the flow will have non-linearities or shock waves, our results implies that
whether or not standing shocks forms will depend on the flow composition.
Thus, for instance, a pure electron-positron pair plasma will never 
undergo a shock transition, while mixing it with some baryons (common in
outflows and jets, for example) as in a completely 
ionized gas, will have shocks. We study in detail how the baryon loading affects 
the shock properties and discuss implication in astrophysical observations.
\end{abstract}


\section{Introduction}	

Accretion is an essential process to explain the electromagnetic radiations coming from microquasars and AGNs. 
The gravitational energy released from the accreting matter is converted into 
kinetic and thermal energies. It is well known that a black hole (\ie BH) has the unique inner boundary condition 
that matter will cross the event horizon at the speed of light $c$ (e.g., Chakrabarti, 1990, hereinafter C90a; 1996ab).
Moreover, the maximum possible sound speed is $a_{\rm max}=c/\sqrt{3}$ and hence close to the BH horizon, 
the accreting matter should be supersonic. As the matter enters the disk sub-sonically and exits from
it to the black hole supersonically, it must pass through a sonic point at least once which makes a black hole 
accretion to be a `transonic' flow. When the flow has some angular momentum, 
the number of physical (saddle type) sonic points could be more than one 
(Chakrabarti, 1989, C90a). This creates a situation that a standing non-linear wave or a shock 
wave may be formed in between these sonic points. The predictions in this
regard have been verified by extensive numerical 
simulations by independent groups and independent numerical
codes (Chakrabarti \& Molteni, 1993; Ryu, Chakrabarti \& Molteni, 
1997; Molteni, Toth \& Kuznetsov, 1999). What is more, the shocks were found to be extremely stable when perturbed
in all three spatial co-ordinates (Molteni, Toth \& Kuznetsov, 1999; Molteni et al., 2001; Nagakura \& Takahashi, 2010).  

The low-viscosity, low angular momentum, transonic flow around a black hole is an indispensable component of black 
hole astrophysics and thus it is very important that we study its properties thoroughly. Together with a high-viscosity, 
non-transonic Keplerian flow (which is known as the standard disk; Shakura \& Sunyaev, 1973;
Novikov \& Thorne, 1973) on the 
equatorial plane, a transonic flow, with or without a shock wave, explains all the observed spectral and timing 
properties of the galactic and extra-galactic black holes (Chakrabarti \& Titarchuk, 1995; Chakrabarti, 
Acharyya \& Molteni, 2004; Wu, Soria \& Campbell-Wilson, 2002; Smith et al. 2001, 2002; 2007; Mandal \& 
Chakrabarti, 2008; Gliozzi et al. 2010; Roy et al. 2011). Oscillations of the shocks in presence of 
cooling is well known (Molteni, Sponholz \& Chakrabarti, 1996; Chakrabarti, Acharyya \& Molteni, 2004)
Moreover, the post-shock region forms the outflows which are observed as the jets. Since the shocks 
are absent in very soft states, the outflows are also found to be absent. 

Since these shocks are centrifugal pressure mediated, doubts persisted whether shocks form in presence of
processes which reduces the angular-momentum of the flow, such as, magnetic field or dissipative processes like viscosity.
There are many numerical as well theoretical investigations of dissipative (\ie in presence of viscosity and/or 
radiative losses), transonic accreting flow around BH which showed the existence of oscillating or standing 
shocks in a significant part of the energy-angular momentum parameter space 
(Chakrabarti, 1996a; Molteni, Sponholz \& Chakrabarti, 1996; Lanzafame, Molteni and Chakrabarti, 1998; 
Chakrabarti \& Das, 2004).
Magnetized flows also show the existence of magneto-critical points and MHD shocks (Chakrabarti, 1990ab;
Takahashi et al. 2006; Das \& Chakrabarti, 2007).

Sonic points in a flow can form only if the equation of state (EoS) of matter is `right'. Even in a 
spherically symmetric adiabatic Bondi flow (Bondi, 1952), the adiabatic index $\Gamma$ must be less than 
$1.67$. Accretion flow which is in vertical equilibrium, $\Gamma$ must be less than $1.5$ in order to have multiple 
sonic points (C90a) and thus shocks are possible only for $\Gamma <1.5$ if the appropriate Rankine-Hugoniot
conditions are satisfied (Landau \& Lifshitz, 1987). It is therefore no surprise that the transonic properties, 
the thermodynamic properties and the consequent radiative properties of the flow will depend strongly on the 
composition. Composition decides heating and cooling processes and thus the instantaneous value of polytropic index. 

It is on this important aspect of the composition dependent transonicity of a relativistic flow
around a black hole, the present paper is devoted to. We will first focus our attention on the thermodynamics 
of this flow. A flow is relativistic if its bulk speed is comparable to $c$ and/or its thermal energy 
is comparable to or greater than its rest energy. So a flow will be in a thermally relativistic domain if
$kT/(mc^2){\gsim}1$ (consequently, adiabatic index $\Gamma \rightarrow 4/3$), and in the non-relativistic domain if
$kT/(mc^2){\ll}1$ ($\Gamma\sim5/3$), where, $T$, $k$, and $m$ are the temperature, the Boltzmann constant and 
the mass of the particles that constitute the flow, respectively. Therefore, a thermally trans-relativistic 
flow cannot be described by fixed $\Gamma$ EoS (Taub, 1948; Ryu, Chattopadhyay \& Choi 2006) 
and relativistic EoS (Chandrasekhar, 1938;  Synge, 1957) has to be employed. Furthermore, 
it is the ratio $T/m$ and not $T$ alone, that 
determines the thermal state of the flow. This implies that the composition of the flow should 
be important. Although there are few instances of the use of relativistic EoS in the study of
the  flow around black holes (Blumenthal and Mathews, 1976; Meliani, Sauty, Tsinganos,
Vlahakis, 2004), but the study of the effect of composition on the flow around BHs were not 
considered. Chattopadhyay \& Ryu
(2009, hereafter CR09), considered a relativistic EoS of multi-species flow composed of
electrons ($e^-$), protons ($p^+$) and positrons ($e^+$) with different proportion
to study the radial flow around black hole. CR09 found that in the temperature range
$10^8 {\rm K} \lsim T \lsim 10^{13}$K, the thermodynamics of the flow is 
significantly different for a flow of different composition. More interestingly, CR09 found
that contrary to the expectation, the transonic electron-positron pair flow (hereafter \ee) is the coolest
and slowest \ie the least relativistic, compared to a flow loaded with protons.
However, the pure electron-proton flow (hereafter, \ep) was not found to be the most relativistic. 
The most relativistic flow was found to be the one in which the proton 
number density is about $20 \%$ of the electron number density. As a result, the density and temperature 
distributions, and consequently the radiation emitted will depend on whether we have a pure electron-positron
plasma or a mixed plasma. It is thus natural to extend the investigations of CR09 to low angular 
momentum flows which are known to have a temperature in between $10^8 {\rm K}$ and $10^{13}$K, 
exactly the range in which the thermodynamics of the relativistic flow depends on the composition.
So far, this study has not been done. We carry out this work and
show that the composition of the flow affects the transonic 
nature of the flow. We present all possible global solutions (connecting the horizon and large 
distances away from the BH) including the shock waves when the composition is varied.
Hence the present effort is to extend the relativistic thermodynamics of CR09 on to the physics of
sub-Keplerian flows (Lu, 1985; Chakrabarti, 1989; C90ab; Kafatos \& Yang, 1994; Yang \& Kafatos, 1995; 
Nobuta \& Hanawa, 1994; Chakrabarti, 1996ab).

The paper is organized as follows, in the next Section, we present
the governing equations. In \S 3, we present the critical point properties and 
the solution procedure. In \S 4, we find the global solutions for the electron-proton flow.
In \S 5, the effects of composition on the solutions of relativistic flow are studied.
Finally, in \S 6, we present discussions and concluding remarks.

\section{Equations of Relativistic Flow}

We assume a fully ionized, non-dissipative, steadily rotating and accreting axisymmetric disk
around a Schwarzschild black hole. The flow is assumed to be 
in hydrostatic equilibrium in a direction perpendicular to the disk plane.
The energy momentum tensor an ideal flow is given by (Wald, 1984),
\begin{eqnarray}
 T^{\mu \nu}=(e+p)u^{\mu}u^{\nu}+pg^{\mu \nu},\label{eq1}
\end{eqnarray}
where, $e$ and $p$ are the energy density and gas pressure respectively, all measured in the local frame. The 
4-velocities are represented by $u^{\mu}$. The equations governing a relativistic flow are given by,
\begin{eqnarray}
T^{\mu \nu}_{; \nu}=0 ~~~~ \mbox{and} ~~~~ (nu^{\nu})_{; \nu}=0,\label{eq2}
\end{eqnarray}
where, $n$ is the particle number density of the flow measured in the local frame.
Henceforth, we use a system of units where $G=M_B=c=1$,
such that the unit of length and time are $r_g=GM_B/c^2$, and $t_g=GM_B/c^3$, respectively.
Here, $G$ and $M_B$ are the universal gravitational constant and the mass of the black hole, respectively.
Under the present set of assumptions, the equation (\ref{eq2}) can be written as the radial component of
relativistic Euler equation, the energy equation and the mass accretion rate equation.
\begin{eqnarray}
\hskip 1.0cm
u^r\frac{du^r}{dr}+\frac{1}{r^2}-(r-3)u^{\phi}u^{\phi}=
-\left(1-\frac{2}{r}+u^ru^r\right)\frac{1}{e+p}
\frac{dp}{dr},\label{eq3}
\end{eqnarray}

\begin{eqnarray}
\frac{de}{dr}-\frac{e+p}{n}\frac{dn}{dr}=0,\label{eq4}
\end{eqnarray}

\begin{eqnarray}
{\dot M}=4\pi r H~\rho~u^r,\label{eq5}
\end{eqnarray}
where, ${\dot M}$ is the mass accretion rate, $H$ is the local height of the disk from 
the equatorial plane, and $\rho$ is the local mass density of the flow. The 4-velocities
along the radial and the azimuthal directions are $u^r$ and $u^{\phi}$.  
The expression of the disk half-height is given by,
\begin{eqnarray}
H{\approx}\sqrt{\frac{p}{\rho}[r^3-\lambda^2(r-2)]},\label{eq6}
\end{eqnarray}
This is an approximate form (Das 2004) of the expression proposed by Lasota \& Abramowicz (1997) 
where the factor $r^3/{\gamma}^2$ has been replaced by
$r^3/{\gamma}^2_{\lambda}$. This approximation introduces errors
for $r<4GM/c^2$ and thus could be safely used in our study.
The conservation equations can be integrated in the form of relativistic Bernoulli equation,
\begin{eqnarray}
{\cal E}=\frac{-(e+p)u_t}{\rho},\label{eq7}
\end{eqnarray}
where, ${\cal E}$ is the Bernoulli parameter or the specific energy of the flow. The specific 
angular momentum $\lambda=-u_{\phi}/u_t$ is also conserved. For dissipative accretion flow, 
viscosity transports $\lambda$ outwards but increases ${\cal E}$ inwards. 
Realistic cooling processes decrease ${\cal E}$ inward. Since due to the stronger 
gravity, the infall time-scale close to the horizon is much shorter than the
dissipation time scales, within few tens of $r_g$, the variation of ${\cal E}$ and $\lambda$
are negligible (Chakrabarti, 1996b; Das, Becker \& Le, 2009). We may thus assume 
that  ${\cal E}$ and $\lambda$ of the flow can be treated as those at the inner boundary 
of a dissipative disk. Consideration of dissipative processes helps
in determining outer boundary conditions (Chakrabarti, 1996b; Lu, Yi and Gu, 2004).

In this paper, we assume the flow to be composed of electrons ($e^-$), protons ($p^+$) and positrons
($e^+$), in varying proportion. The total number density is given by,
\begin{eqnarray}
n={\Sigma} n_j=n_{e^-}+n_{e^+}+n_{p^+}=2n_{e^-},\mbox{and} ~~~~ n_{e^+}=n_{e^-}(1-\xi),\label{eq8}
\end{eqnarray}
where, the composition parameter $\xi=n_{p^+}/n_{e^-}$ is the ratio of proton and electron number 
densities (\eg CR09). Hence, the electron-positron flow (\ie $e^--e^+$) is described 
by $\xi=0$, similarly the electron-proton flow ($e^--p^+$) is described by $\xi=1$.
The mass-density is given by,
\begin{eqnarray}
\rho=\Sigma n_jm_j=n_{e^-}m_{e^-}\left\{ 2-\xi \left(1-\frac{1}{\eta}\right)\right\},\label{eq9}
\end{eqnarray}
where, $\eta=m_{e^-}/m_{p^+}$, $m_{e^-}$ and $m_{p^+}$ are the electron and proton mass respectively.
The isotropic pressure and energy density for single temperature flow is given by (CR09)
\begin{eqnarray}
p=\Sigma p_j=2n_{e^-}kT, ~ \mbox{and} ~~~~ e=\Sigma e_j=n_{e^-}m_{e^-}f,\label{eq10}
\end{eqnarray}
where,
\begin{eqnarray}
f=(2-\xi)\left[1+\Theta\left(\frac{9\Theta+3}{3\Theta+2}\right)\right]
+\xi\left[\frac{1}{\eta}+\Theta\left(\frac{9\Theta+3/\eta}{3\Theta+2/\eta}\right)\right],\label{eq11}
\end{eqnarray}
and, $\Theta=kT/m_{e^-}$. The expression of polytropic index, adiabatic index and the sound speed (CR09)
are given by
\begin{eqnarray}
N=\frac{1}{2}\frac{df}{d\Theta}, ~~~~ \Gamma=1+\frac{1}{N}, ~~~~ \mbox{and} ~~~~ a^2=\frac{2 \Gamma \Theta}{f+2\Theta}.\label{eq12}
\end{eqnarray}
It has been shown in CR09 that $\Gamma$ remains roughly constant at $5/3$ for  $T \lsim10^8$K, 
while $\Gamma \rightarrow 4/3$ and remains constant for $T\gsim10^{13}$K, independent of the 
value of $\xi$. In the temperature range $10^8$K $<T<10^{13}$K,
$\Gamma$ (or $N$) depends both on $T$ and $\xi$.

\section{Critical point analysis}

Equations (\ref{eq3}-\ref{eq5}), with the help of equations (\ref{eq6}, \ref{eq8}-\ref{eq12}), 
can be simplified to,
\begin{eqnarray}
\frac{dv}{dr}=\frac{{\cal G}a^2{\cal A}_1+{\cal A}_2-{\cal A}_3}{\gamma^2_v[v-2{\cal G}a^2/v]},\label{eq13}
\end{eqnarray}
and
\begin{eqnarray}
\frac{d\Theta}{dr}=-\frac{2 {\cal G}\Theta}{N}\left[\frac{r-1}{r(r-2)}+\frac{{\cal A}_4}{2}+\frac{1}{v(1-v^2)}
\frac{dv}{dr}\right],\label{eq14}
\end{eqnarray}
while, various expressions in the above two equations are given by
\begin{eqnarray}
{\cal A}_1=\frac{2(r-1)}{r(r-2)}+{\cal A}_4; ~~~~ {\cal A}_2=\frac{(r-3)\lambda^2\gamma^2_{\lambda}}{r^4}  \nonumber \\
 {\cal A}_3=\frac{1}{r(r-2)}; ~~~~ {\cal A}_4=\frac{(3r^2-\lambda^2)}{r^3-\lambda^2(r-2)} 
\nonumber \\
 {\cal G}=\frac{N}{2N+1}; ~~~~ \gamma^2_{\lambda}=\frac{r^3}{r^3-\lambda^2(r-2)}; 
\nonumber \\
 \mbox{and} ~~~~ \gamma^2_v=\frac{1}{1-v^2}\label{eq15}
\end{eqnarray}

In equations (\ref{eq13}-\ref{eq14}), the radial three velocity in co-rotating frame 
is defined as $v^2=\gamma^2_{\lambda}v^2_s$, where $v^2_s=-u_ru^r/(u_tu^t)$.  
At $r=r_c$, the critical point of the flow, $dv/dr={\cal N}/{\cal D}{\longrightarrow}0/0$
which gives us the so called critical point conditions (Chakrabarti, 1990a)
\begin{eqnarray}
v_c=(2{\cal G}_c)^{1/2}a_c ~~~ {\rm and} ~~~  
a^2_c=\frac{1/[r_c(r_c-2)]-\lambda^2(r_c-3)\gamma^2_{{\lambda}c}/r^4_c}{{\cal G}_c{\cal A}_{1c}},
\label{eq16}
\end{eqnarray}
where, any quantity $Q$ evaluated at $r_c$ has been presented as $Q_c$.
The gradient of velocity at $r_c$ is obtained by l'Hospital rule,
\begin{eqnarray}
\left(\frac{dv}{dr}\right)_{r_c}=\frac{(d{\cal N}/dr)_{r_c}}{(d{\cal D}/dr)_{r_c}},
\label{eq17}
\end{eqnarray}
where, ${\cal N}$ is the numerator and ${\cal D}$ is the denominator of $dv/dr$.
Equation (\ref{eq17}) is a quadratic equation with two roots and the 
solution with the negative root is the accretion branch.
The nature of the critical point is obtained by calculating $\left(dM/dr\right)_{r_c}$ ($M=v/a$ is the Mach number) 
(Chakrabarti, 1990a). From equation (\ref{eq14}), we also get,
\begin{eqnarray}
\left(\frac{d\Theta}{dr}\right)_{r_c}=-\frac{2 {\cal G}_c\Theta_c}{N_c}\left[\frac{r_c-1}{r_c(r_c-2)}+\frac{{\cal A}_{4c}}{2}+\frac{1}{v_c(1-v^2_c)}
\left(\frac{dv}{dr}\right)_{r_c}\right].
\label{eq18}
\end{eqnarray}

Presently, we are only concentrating on non-dissipative processes, and no particle creation/annihilation 
etc. are considered. Therefore, ${\cal E}$, $\lambda$ and $\xi$ are constants along the flow.

\begin{figure} [h]
\begin{center}
{
\includegraphics[scale=0.43]{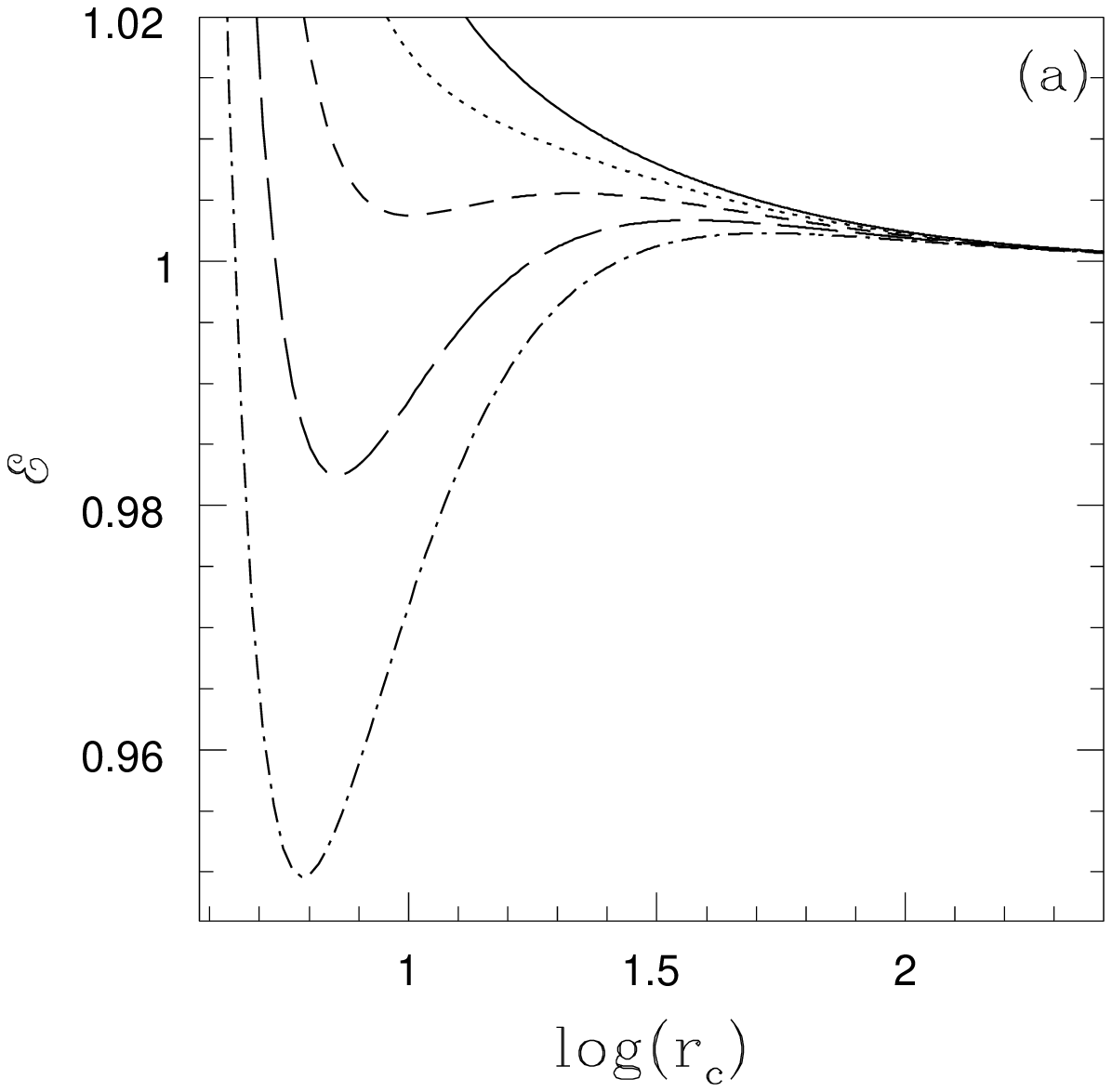}\\
\includegraphics[scale=0.43]{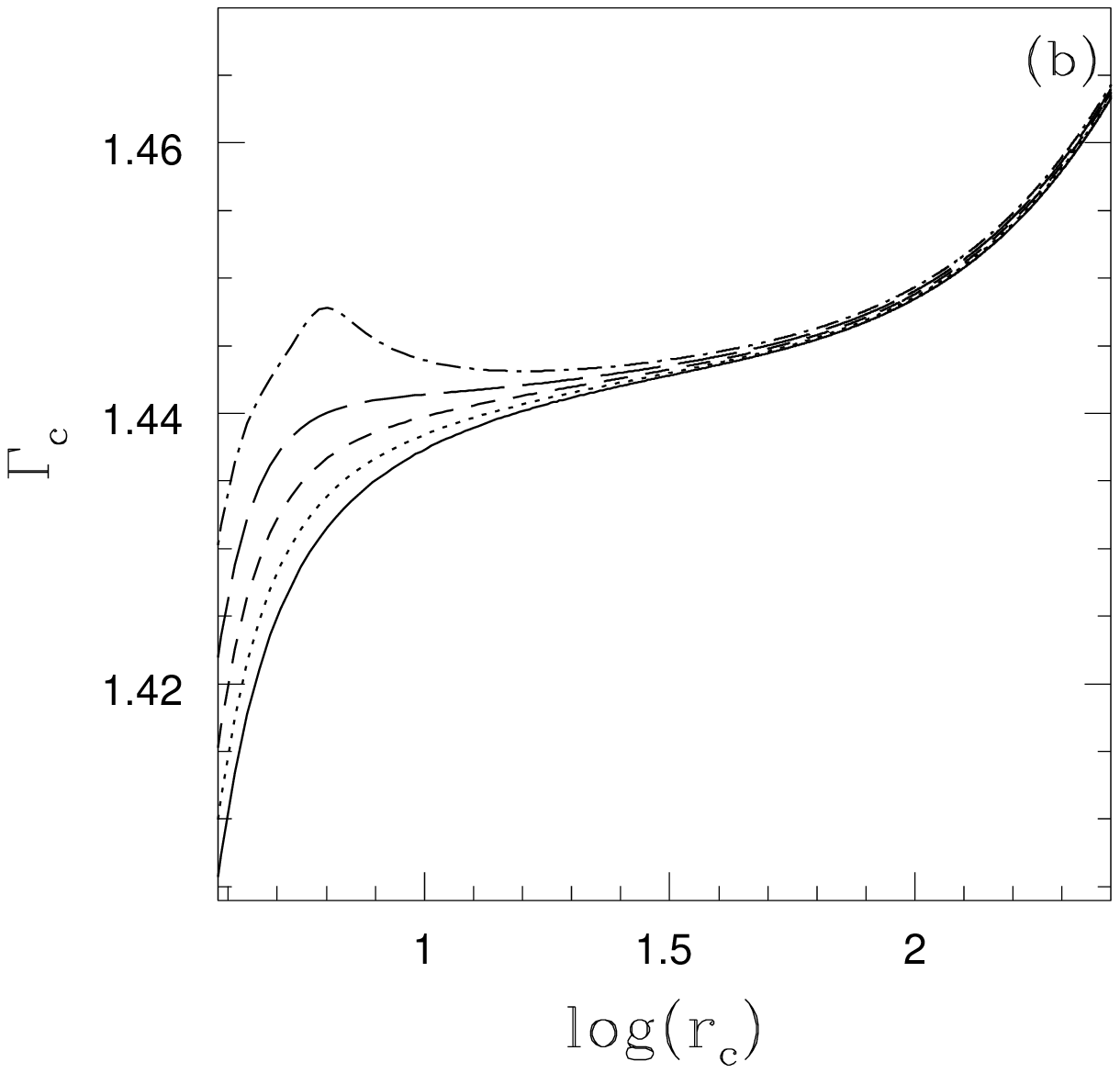}}
\end{center}
\caption{Variation of (a) ${\cal E}$ and (b) ${\Gamma}_c$ with $log(r_c)$, for $\lambda=2$ (solid),
$2.4$ (dotted), $2.8$ (dashed), $3.2$ (long-dashed), $3.6$ (dashed-dot). Both the plots are for \ep flow.}
\label{fig1}
\end{figure}

The procedure to obtain a complete solution is the following ---
equations (\ref{eq12}) and (\ref{eq16}) are combined to express $\Theta_c$ in terms of $r_c$,
$\lambda$ and $\xi$. Using this $\Theta_c$ in equations (\ref{eq7}) and (\ref{eq11}), we
get a formula involving ${\cal E}$, $r_c$,
$\lambda$ and $\xi$. For a given value of ${\cal E}$, $\lambda$ and $\xi$, $r_c$ is computed from the formula. In significant part of the
${\cal E}$-$\lambda$ space, {\it three roots} for $r_c$ are found, which are the inner ($r_{ci}$), the middle ($r_{cm}$)
and the outer ($r_{co}$) critical points. Moreover, $r_{ci}$ and $r_{co}$ are the physical saddle type critical points, while $r_{cm}$ is unphysical centre type.
We consider $r_c$ (either $r_{ci}$ or $r_{co}$ or, both if they exist), and the corresponding $a_c$, $v_c$,
$(dv/dr)_{r_c}$ and $\left(d\Theta/dr\right)_{r_c}$ as the initial values
and integrate equations (\ref{eq13}-\ref{eq14}) once inward and then outward, to obtain the solutions.

\begin{figure}
\begin{center}
\includegraphics[scale=0.64]{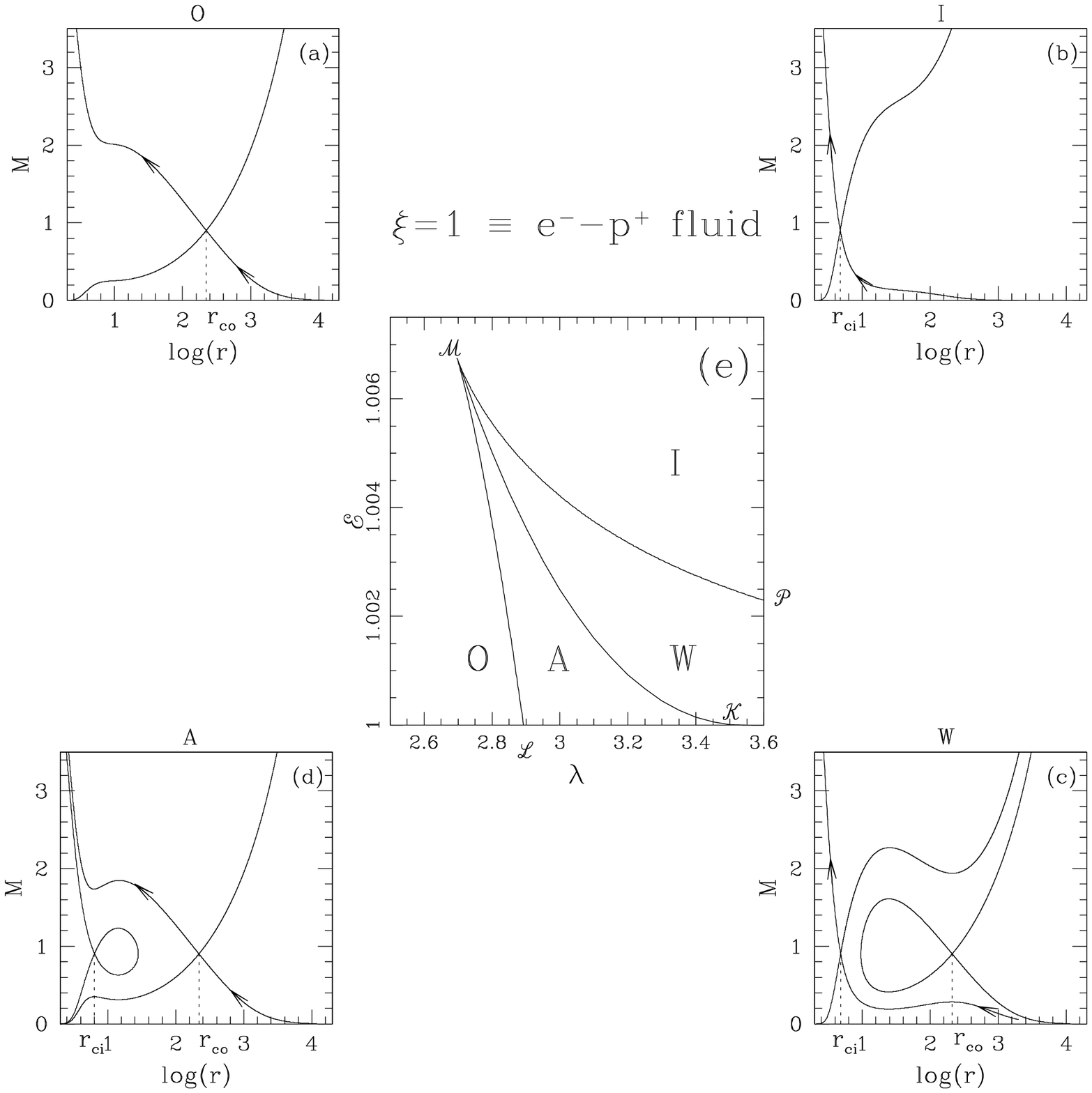}
\end{center}
\caption{The domain for multiple-critical point in ${\cal E}- \lambda$ space, is MCP region (e). $M-log(r_c)$ plot
of the O type (a); I type (b); W type (c); and A type (d), solutions are presented. The solution type are also marked
above each Figure. The arrows mark the smooth global accretion solutions. All the Figures are for 
\ep flow ($\xi=1.0$). The dotted vertical lines mark the positions of physical critical points.}
\label{fig2}
\end{figure}

\section{Global Solutions of Electron-Proton flow}

We vary two conserved quantities, namely, ${\cal E}$ and $\lambda$  and  present all possible global solutions.
As a representative case, we consider \ep flow (\ie $\xi=1$) in this section.
We plot ${\cal E}$ (Fig. 1a), and ${\Gamma}_c$ (Fig. 1b) with $log(r_c)$, where each curve is for $\lambda=2.0$ (solid),
$2.4$ (dotted), $2.8$ (dashed), $3.2$ (long-dashed), $3.6$ (dashed-dot), respectively.
Since, ${\Gamma}_c$ depends on both $r_c$ and $\lambda$, therefore, {\it for the set}
\{${\cal E}$,${\lambda}$\} which admits multiple critical points, ${\Gamma}_c$ will be different at
$r_{ci}$ {\it and} $r_{co}$. In Figs. 2(a-e) we show various topologies of our solution and the 
division of the parameter space spanned by ${\cal E}$ and ${\lambda}$.
In Fig. 2e, the domain for multiple critical points is given by the region under 
the curve ${\cal LMP}$ (hereafter, this will be referred to as Multi-Critical-Point region,
or MCP region for short.). This curve was obtained by joining the extrema of the curves of Fig. 1a.
In Figs. 2(a-d), arrows mark the smooth global accretion solutions.
The solution types are given bellow: \\
Type O: Solution for any set \{${\cal E},\lambda$\} from the region O
of Fig. 2e, and possess only the $r_{co}$. A sample
solution of $M$ with $log(r)$ is given in Fig. 2a for $\{{\cal E},\lambda\}=\{1.00087,2.8\}$.  \\
Type I: Figure 2b presents a sample solution for $\{{\cal E},\lambda\}=\{1.0055,3.4\}$ from the region I of Fig. 2e, which possess only the $r_{ci}$. \\
Type W: Solution from the region W of Fig. 2e. A sample
solution for $\{{\cal E},\lambda\}=\{1.00087,3.4\}$ is presented in Fig. 2c. Global solution is through $r_{ci}$, and $\alpha$ type through $r_{co}$. \\
Type A: Solution from the region A of Fig. 2e. A sample
solution for $\{{\cal E},\lambda\}=\{1.00087,3.0\}$ is presented in Fig. 2d. The smooth global solution is through $r_{co}$, and reflected-$\alpha$ type through $r_{ci}$. \\
The type of solution is marked above each Figure.

\begin{figure}[h]
\begin{center}
{\includegraphics[scale=0.4]{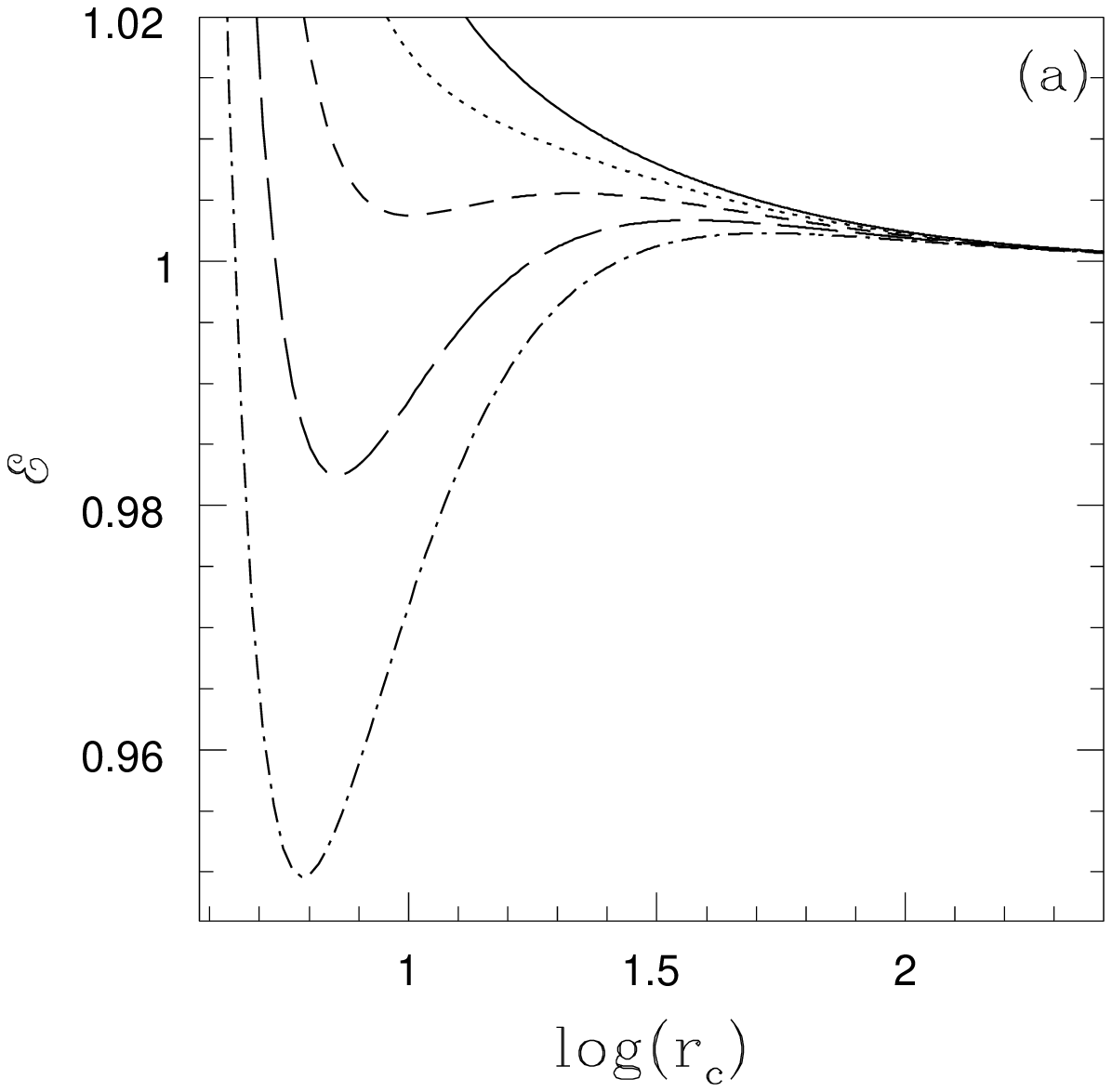}
\vskip -2.0cm
\includegraphics[scale=0.4]{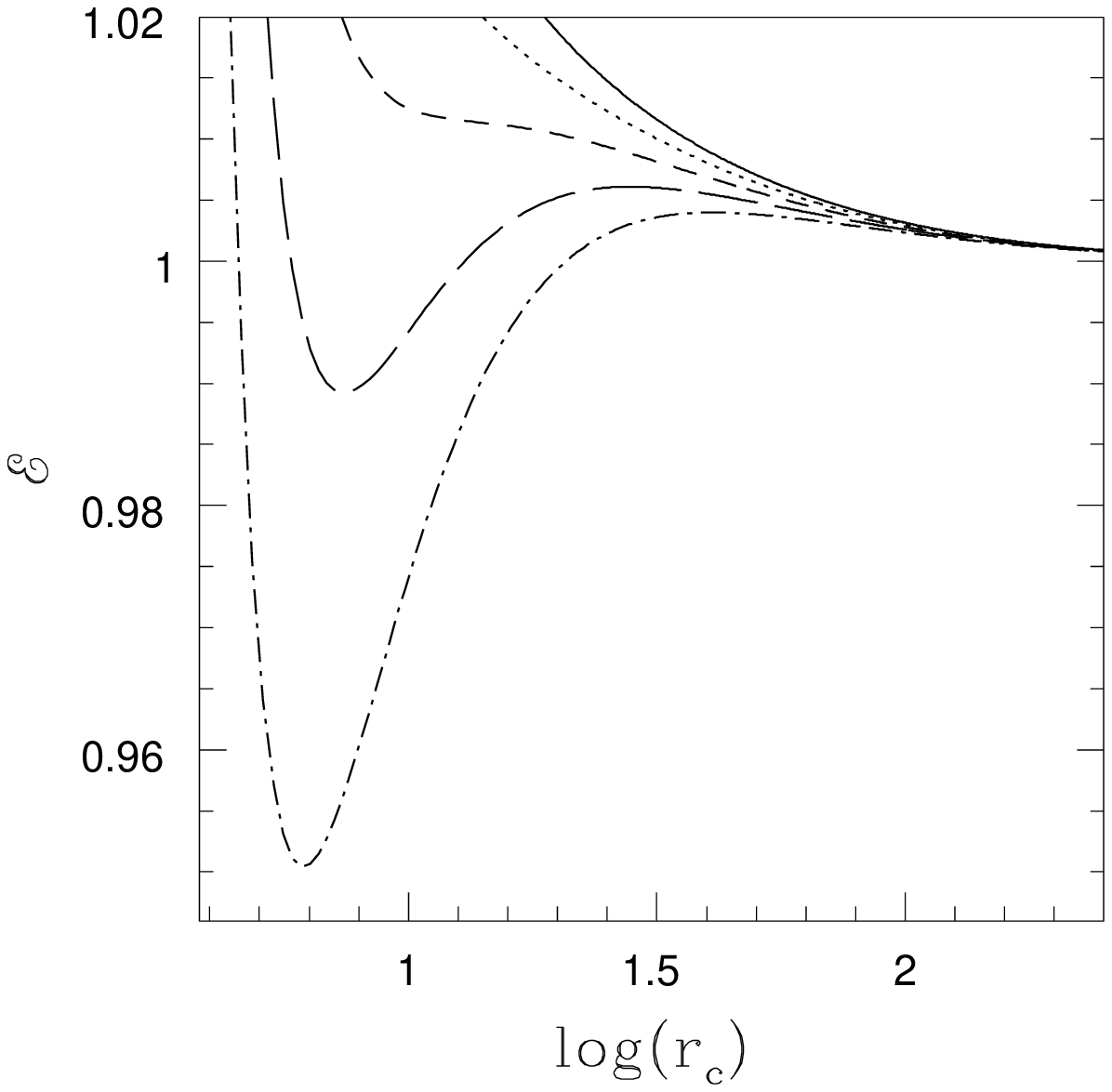}
\vskip -2.0cm
\includegraphics[scale=0.4]{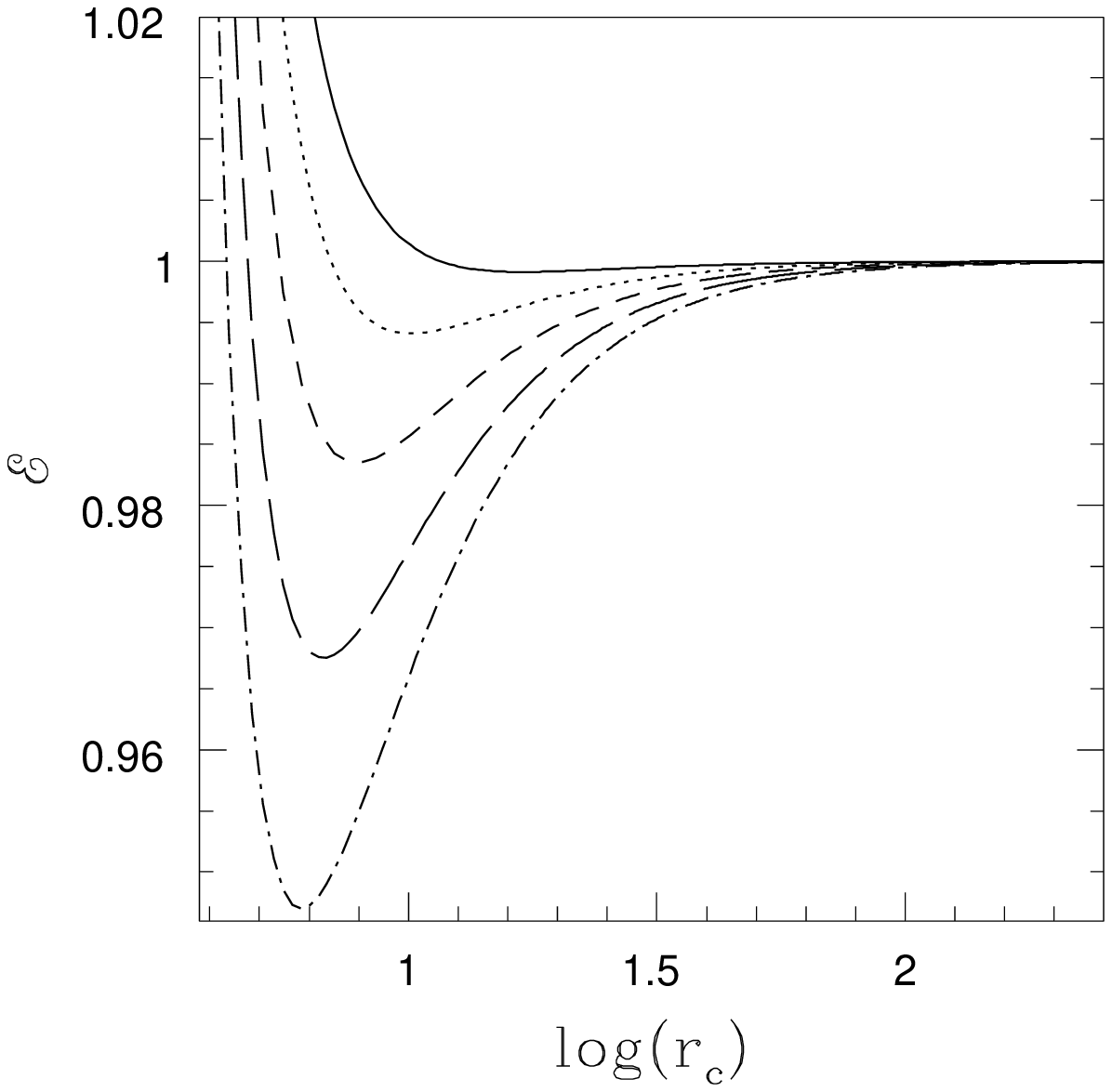}}
\end{center}
\vskip -1.5cm
\caption{${\cal E}$ with $log(r_c)$ for $\lambda=2$ (solid), $2.4$ (dotted), $2.8$ (dashed),
$3.2$ (long-dashed), $3.6$ (dashed-dotted). Each Figure is characterized by $\xi=1.0$ or \ep flow (a), $\xi=0.5$
(b) and $\xi=0.0$ or \ee flow (c).} 
\label{fig3}
\end{figure}

\begin{figure}%

\begin{center}
\includegraphics[scale=0.6]{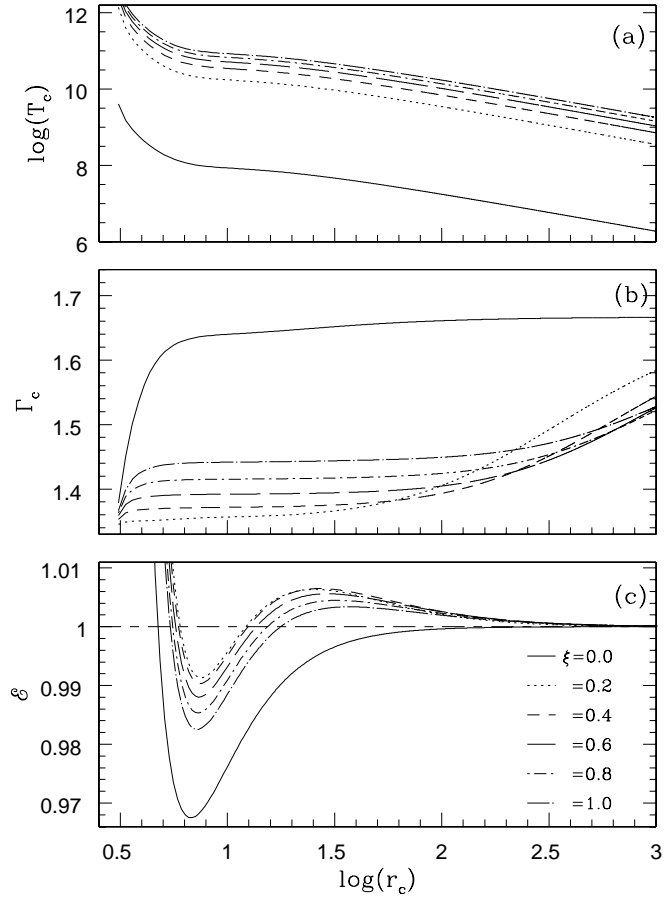}
\end{center}

\caption{Variation of (a)  $log(T_c)$, (b) $\Gamma_c$ and (c) ${\cal E}$ at the critical points with $log(r_c)$, 
for $\xi=0.0$ (solid), $0.2$ (dotted), $0.4$ (dashed), $0.6$ (long dashed), $0.8$ (dashed-dotted) and $1.0$ 
(long dashed-dotted). The long-short dashed curve in the bottom panel indicates the bound energy level. 
All the curves are drawn for $\lambda=3.2$.}
\label{fig4}
\end{figure}

\begin{figure}[h]
\begin{center}
{
\includegraphics[scale=0.35]{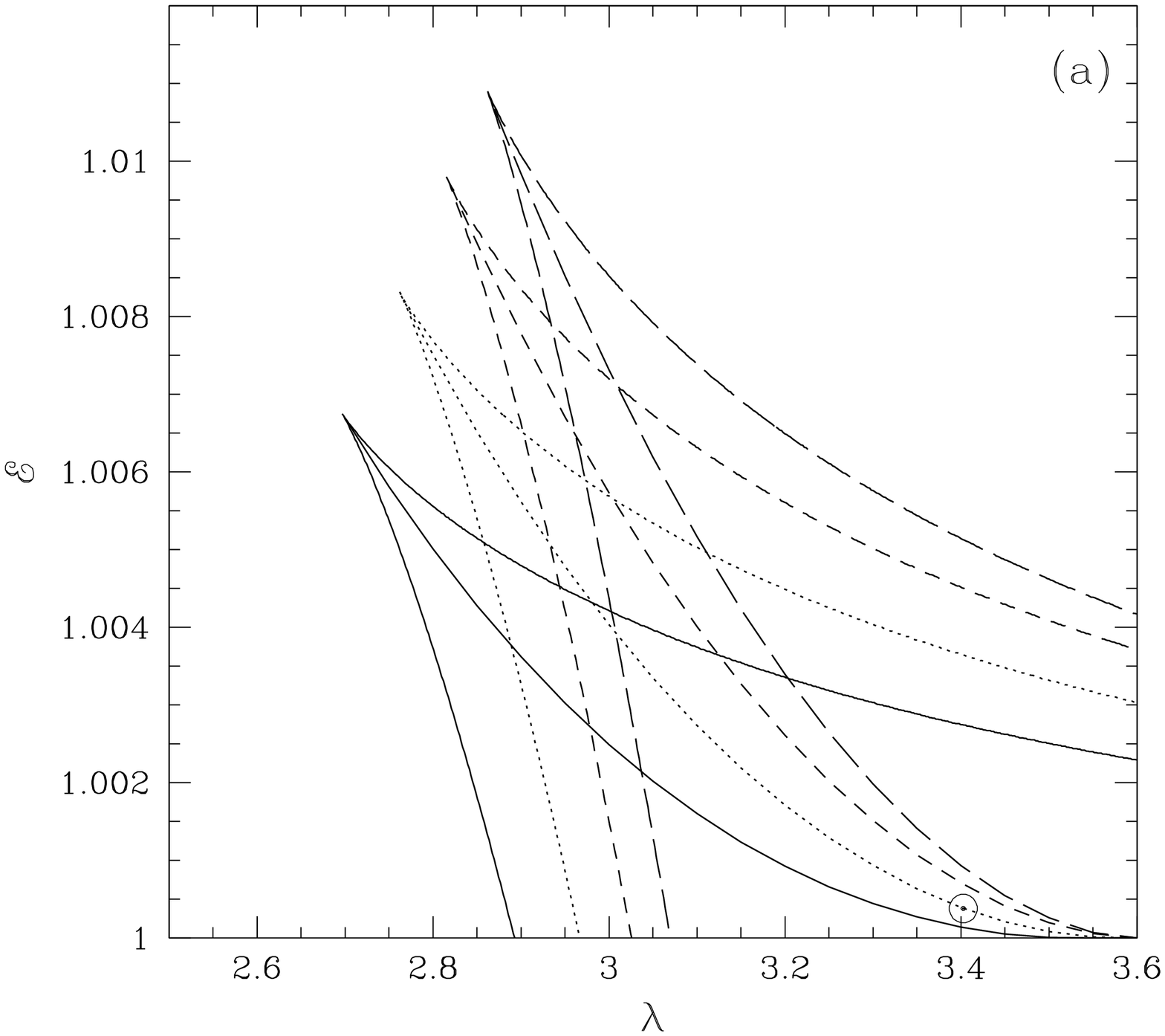}
\vskip -1.0cm
\includegraphics[scale=0.35]{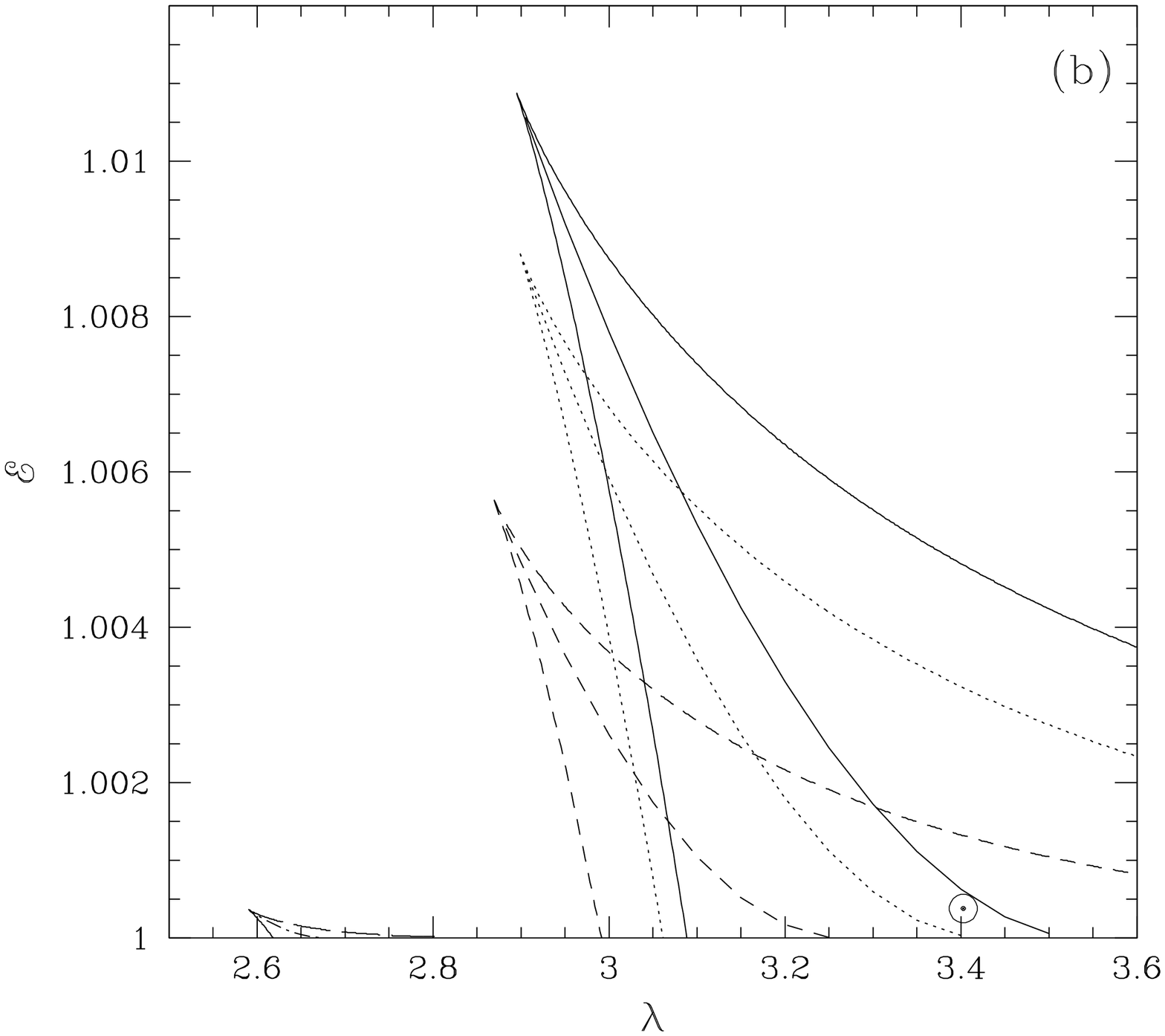}
}
\end{center}
\caption{(a) The ${\cal E}-\lambda$ parameter space for multiple critical point 
(the MCP region) for $\xi=1$ (solid), $0.8$ (dotted), $0.6$ (dashed), 
$0.4$ (long dashed). (b) The MCP region for $\xi=0.2$ (solid), $0.1$ (dotted), 
$0.05$ (dashed), $0.01$ (long dashed). The circled-dot is located at ${\cal E}=1.0004$ 
and $\lambda=3.4$ of the parameter space. MCP region is not explicitly mentioned
to avoid clumsiness.}
\label{fig5}
\end{figure}

\begin{figure}[h]
\begin{center}
{
\includegraphics[scale=0.35]{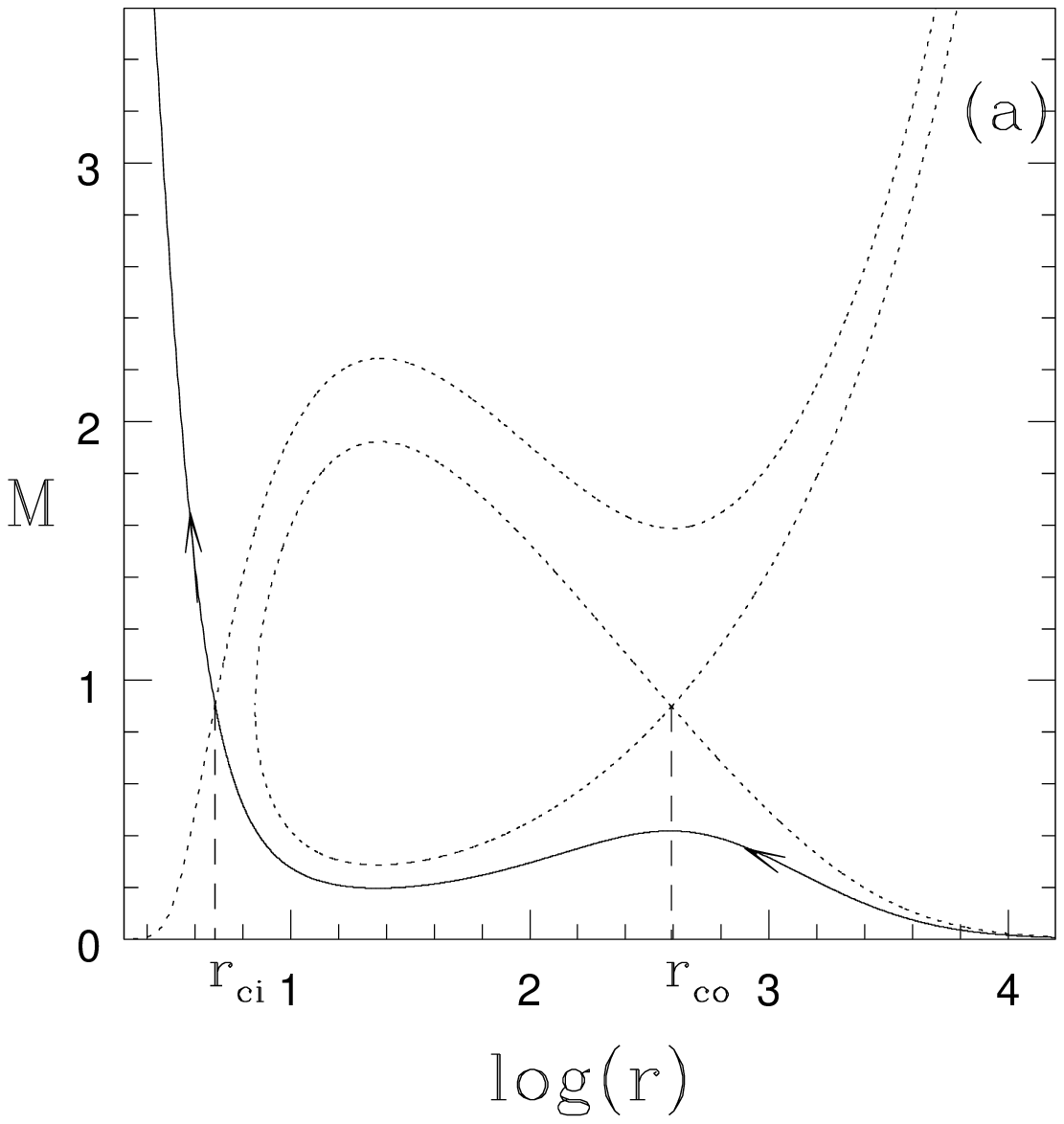}
\vskip -1.0cm
\includegraphics[scale=0.35]{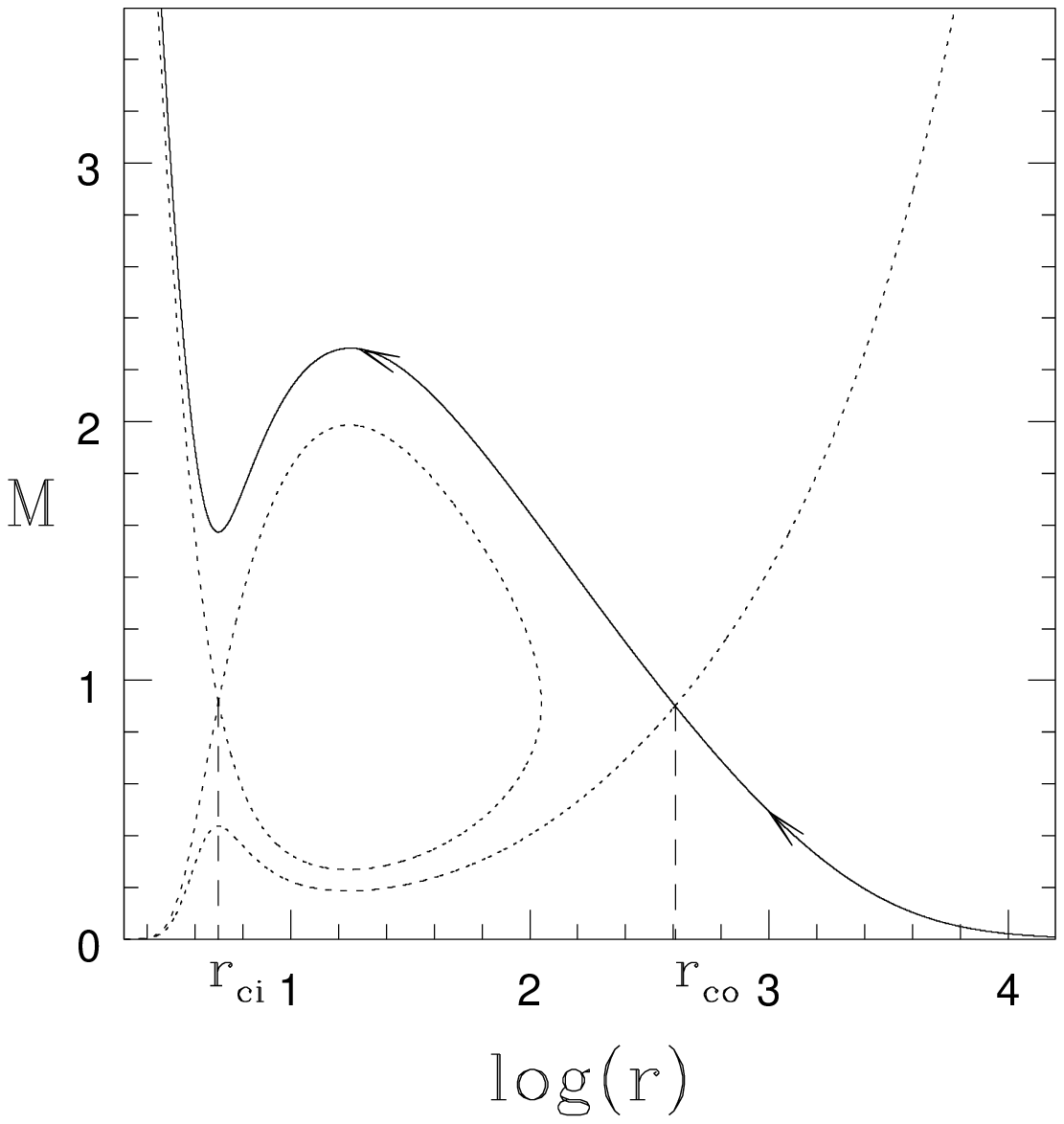}
\vskip -1.0cm
\includegraphics[scale=0.35]{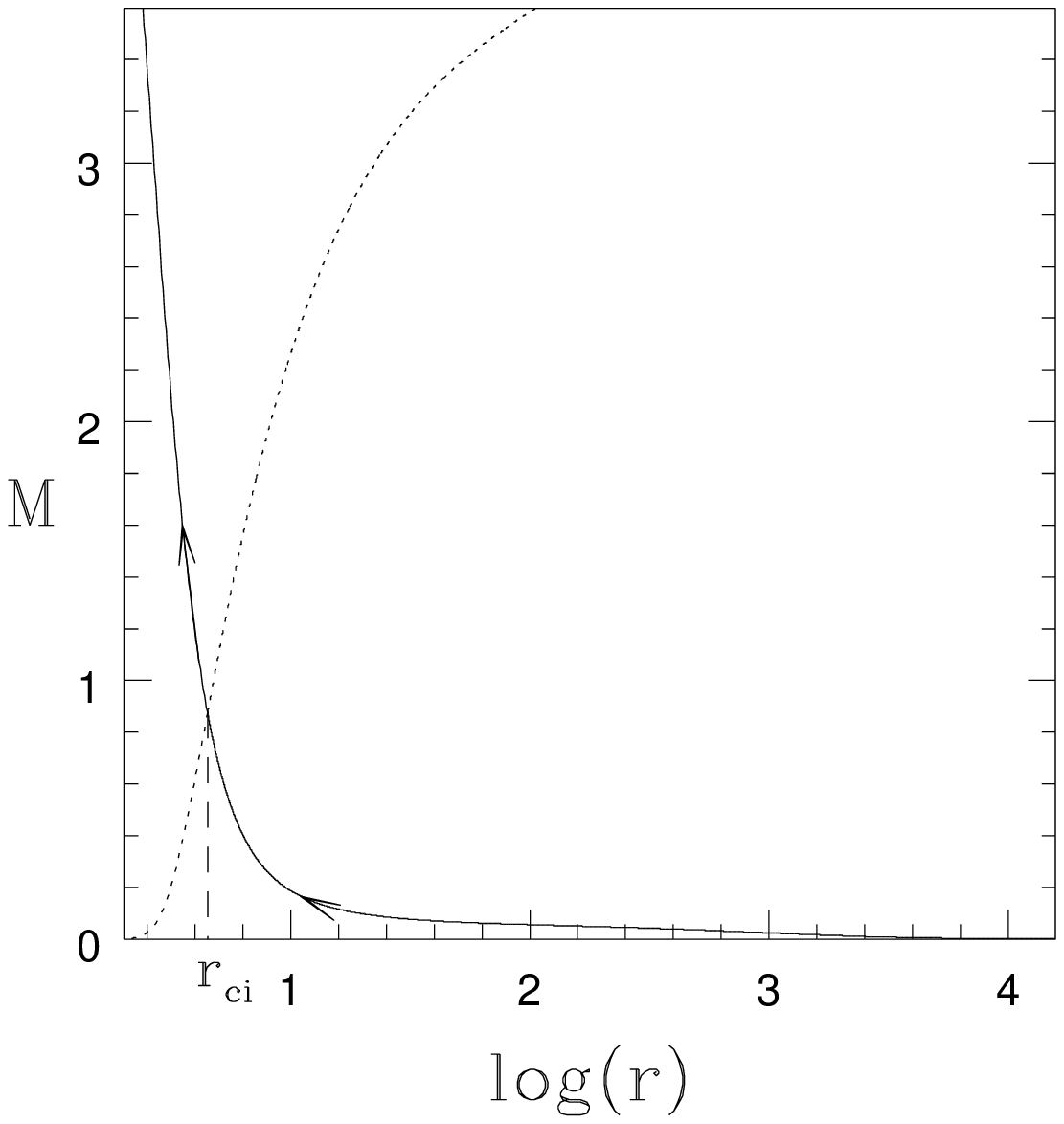}

}
\end{center}
\vskip -1.5cm
\caption{The Mach number $M$ is plotted with $log(r)$ for $\{{\cal E},\lambda\}=$\{$1.0004,3.4$\}. (a) \ep 
flow \ie $\xi=1.0$, the dotted curve through $r_{ci}$ is the wind type solution
and the $\alpha$ type solution is through $r_{co}$. (b) A flow with $\xi=0.5$, the dotted curve through $r_{co}$
is the wind type solution and the reflected-$\alpha$ type solution is through $r_{ci}$. 
(c) \ee flow \ie $\xi=0.0$ and the dotted curve through $r_{ci}$ is a wind type solution. In all the figures,
the solid curve with arrows, are the smooth global accretion solutions.}
\label{fig6}
\end{figure}

\section{The effect of composition of the flow}

We compare the ${\cal E}-log(r_c)$ for $\xi=1.0$ (Fig. 3a), $\xi=0.5$ (Fig. 3b)
and $\xi=0.0$ (Fig. 3c). Each curve is for $\lambda=2.0$ (solid), $2.4$ (dotted), $2.8$ (dashed), $3.2$ (long-dashed), $3.6$ (dashed-dotted).
In Fig. 3a, ${\cal E}$ monotonically decreases with $r_c$ for lower $\lambda$ ($=2,~2.4$), and hence
for these values of $\lambda$ there is only single critical point. For higher $\lambda$ ($=2.8,~3.2,~3.6$), each of the ${\cal E}$ curves has a pair of
maximum and minimum and therefore multiple critical points do form. In Fig. 3b, higher $\lambda$ is required to form multiple critical points (\ie for $\lambda \gsim 2.8389$).
However, the most spectacular
difference is in Fig. 3c!
In Fig. 3c, for any value of $\lambda$, each of the ${\cal E}-log(r_c)$ curves have only a minimum and no
maximum.
Once the ${\cal E}$ curve for a particular $\lambda$ becomes bound (\ie ${\cal E}<1$), it never manages to be greater than one but only approaches ${\cal E}(r_c){\rightarrow}1$ asymptotically (\ie as $r_c{\rightarrow}{\infty}$). This implies that, {\it in the case of} \ee
{\it flow, there is only one} $r_c$ when ${\cal E}>1$, independent of the value of
$\lambda$. There are two $r_c$ when ${\cal E}<1$, but only the inner one is saddle type and the outer is
centre type.

In order to get a clearer idea of the effect of $\xi$, we plot $log(T_c)$ 
(Fig. 4a), ${\Gamma}_c$ (Fig. 4b) and ${\cal E}$ (Fig. 4c) with $log(r_c)$.
Each curve is for $\xi=0.0$ (solid), $0.2$ (dotted), $0.4$ (dashed), $0.6$ (long dashed),
$0.8$ (dashed-dotted) and $1.0$ (long dashed-dotted), and $\lambda=3.2$ is same for all the curves.
Figure 4a shows that $T_c$ increases with $\xi$ and hence \ee has the lowest $T_c$.
From equations (\ref{eq16}) it is clear that $v_c$ explicitly depends on $r_c$ and $\lambda$ 
and implicitly on $\xi$. It is obvious that $a_c$ should depend on $\xi$ 
because of the presence of $N_c$ in its expression. However, since $N_c$ is present in
both the numerator and denominator of ${\cal G}_c$ [see equation (\ref{eq15})], the dependence of $a_c$ on $\xi$ 
is very weak. So, $a_c$ strongly depends on $r_c$ and $\lambda$ but very weakly on $\xi$. Therefore, 
at a given $\lambda$, plotting any quantity as a function of $r_c$ is equivalent to plotting 
them as a function of $a_c$. 

For a given sound speed, the momentum transferred by the lighter particles in the flow frame, is far less 
than the heavier particles. As a result, $T_c$ of \ee flow is much less compared to that of the flow 
with finite baryons (see, Figs. 1a-1c of CR09). Thus at a given $r_c$, increasing the fraction of proton
increases the temperature of the transonic flow (Fig. 4a). But by the same token the rest energy 
(per oppositely charged particles) increases too.
We know that the thermal state of a flow is determined by the competition between the thermal and the
rest energy, and hence beyond a limit, addition of protons would make the 
flow hotter but thermally less relativistic. Hence Fig. 4b shows that $\Gamma_c$ decreases 
(thermally more relativistic) from $\xi=0.0$ up to a certain value, say $\xi=\xi_{\rm lim}$, 
and then increases (thermally less
relativistic) for higher $\xi$. In Figure 1c of CR09, it was shown 
that $\xi_{\rm lim}$ is higher for lower values of $a$ and vice versa. 
Similarly, $\xi_{\rm lim}$ will depend on $r_c$ too, for \eg
$\xi_{lim}\sim 0.24$ for $3<r_c<54$ (\ie for $0.4>a_c>0.09$), while $\xi_{\rm lim}\sim 0.4$ for $r_c\sim 100$.
The overall similarity of Fig. 1c of CR09 and Fig. 4b is quite obvious.
In Fig. 4c, ${\cal E}(r_c)$ increases with $\xi$ up to $\xi_{\rm lim}=0.24$ within the range $3<r_c<54$. 
But the more interesting thing to note is the effect of $\xi$ on the formation of multiple critical points. 
For \ee flow, ${\cal E}-log(r_c)$ curve shows that there is only one minimum 
independent of the value of $\lambda$. Since relativistic temperatures are achieved closer to the horizon
for \ee flow, so physical critical points are also achieved closer to the horizon. 
On the other hand, for a flow with $1 \geq \xi > 0$ there are both
maxima and minima, and so, there may be more than one physical critical point if ${\cal E}>1$.


Similar to Fig. 2e, the MCP regions of ${\cal E}-\lambda$ space that admits multiple critical points 
can be found out by joining the extrema of ${\cal E}$---$log(r_c)$ curves for 
flow of respective $\xi$. In Fig. 5a, the MCP region is compared for $\xi=1$ (solid), $0.8$ (dotted), $0.6$
(dashed), $0.4$ (long dashed). In Fig. 5b, the MCP region for
$\xi=0.2$ (solid), $0.1$ (dotted), $0.05$ (dashed), $0.01$ (long dashed) are plotted. 
The MCP region shifts rightwards (to higher ${\cal E}$ and 
$\lambda$) when $\xi$ is decreased in the range $1.0 \geq \xi \gsim 0.24$, but
moves to the left (to lesser ${\cal E}$ and $\lambda$) when $\xi$ is decreased 
in the range $0.24 > \xi > 0$. The MCP domain for $\xi=0.01$ (long-dashed in Fig. 5b) is almost 
imperceptible, and it actually vanishes for $\xi{\sim}0$. The closed region MCP do not form for \ee flow,
because there is no maxima in the ${\cal E}-log(r_c)$ plot -- only minima are present.
Therefore, the response of MCP region to the change in $\xi$ is not monotonic.
Figures 5(a-b) imply that not only transonic-rotating \ee flows are 
quantitatively less energetic and less relativistic than flow of non-negligible protons, 
but qualitatively differ since multiple critical points are not found for any $\lambda$. 

Furthermore, from Figs. 5(a-b) it is clear that an O type domain for a 
particular $\xi$, may be a domain for A type or W type or I type for a different $\xi$. 
To illustrate this, we choose a set $\{{\cal E},\lambda\}=\{1.0004,3.4\}$ (marked circled-dot in Figs. 5a-5b)
and plot $M-log(r)$ for $\xi=1$ (Fig. 6a), $\xi=0.5$ (Fig. 6b) and $\xi=0$
(Fig. 6c). Clearly, for the same set of $\{{\cal E},\lambda\}$ the \ep flow is W type, while for $\xi=0.5$ and \ee flow the solutions are A type and I type, respectively. Thus, the difference in solutions with the change of $\xi$, is not just {\it quantitative}, but {\it qualitative} as well.

\begin{figure}[h]

\begin{center}
{
\includegraphics[scale=0.8]{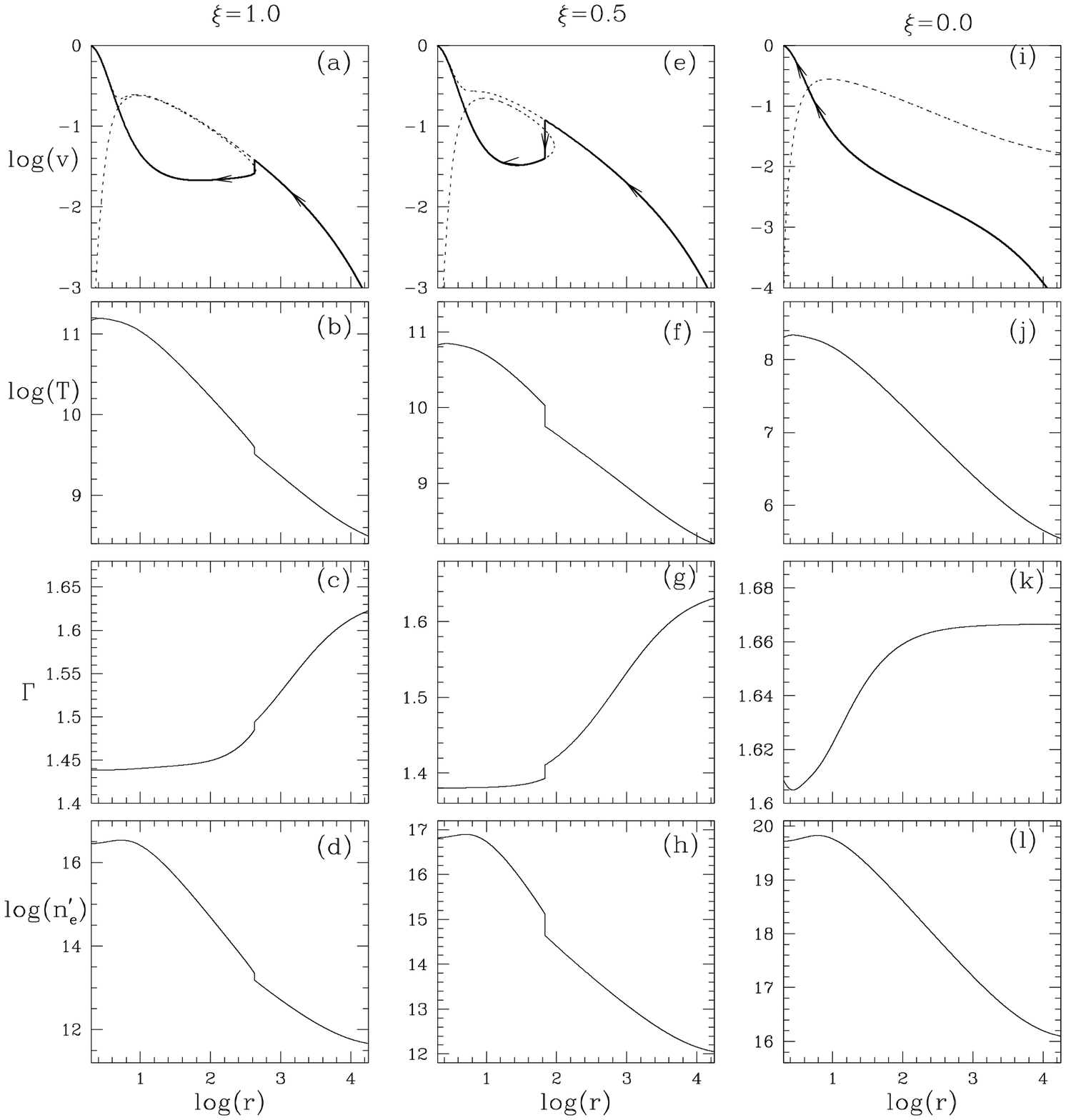}
}
\end{center}

\caption{Various flow variables are plotted for $\{{\cal E},\lambda\}=\{1.00009,3.4\}$.
The first column is for $\xi=1.0$ (Figs: a,b,c,d); the second for $\xi=0.5$ (Figs: e,f,g,h)
and the third for $\xi=0.0$ (Figs: i,j,k,l), respectively. Each row represents flow variables $log(v)$
(Figs: a,e,i);
$log(T)$ (Figs: b,f,j); $\Gamma$ (Figs: c,g,k) and $log(n^{\prime}_e)$ (Figs: d,h,l); where $n^{\prime}_e=n_{e^+}+
n_{e^-}$.
Vertical jumps represent standing shocks. For $\xi=1.0$ $x_s=423.46$, and for $\xi=0.5$ it is $x_s=68.78$. Solid curve
with arrow, on the top row represents solutions chosen by the accreting matter out of all relevant solutions (dotted).}
\label{fig7}
\end{figure}

\subsection{Accretion shocks}
One of the major consequences of the existence of multiple physical $r_c$, is the 
possibility of the formation of shocks in the accretion disk around a black hole 
(Chakrabarti, 1989; 1996ab; Takahashi et al. 2006; Fukumura \& Kazanas, 2007; Das, Becker
and Le 2009). At $r{\sim}$few$\times 10r_g$, a flow accreting through $r_{co}$ is slowed down by the 
twin effect of increasing gas pressure and the centrifugal term.
If this effect is strong enough, then the slowed down region of the flow acts as the effective 
boundary layer to the faster flow following behind, and causes formation of a shock wave
in an accretion flow.

The relativistic Rankine-Hugoniot conditions which must be satisfied at the shock are 
(Taub, 1948; Chakrabarti, 1996ab, Lu et al. 1997),
\begin{eqnarray}
  [nu^r]=0, \label{eq19}
\end{eqnarray}
\begin{eqnarray}
   [(e+p)u^tu^r]=0, \label{eq20}
\end{eqnarray}
\begin{eqnarray}
[(e+p)u^ru^r+pg^{rr}]=0,\label{eq21}
\end{eqnarray}
where, the square brackets denote the difference of quantities across the shock. Since the information of $\xi$ is 
in the Rankine-Hugoniot conditions, therefore the shock location and its strength will be influenced 
by $\xi$ too. Equations (\ref{eq19}-\ref{eq21}) are checked along the flow to find the shock location $x_s$.
In Figs. 7a-7l, a variety of flow variables are plotted for $\xi=1.0$ (left panels or Figs. 7a-7d), 
$\xi=0.5$ (middle panels or Figs. 7e-7h), and $\xi=0.0$ (right panels or Figs. 7i-7l) 
for the same set of $\{{\cal E},\lambda\}=\{1.00009,3.4\}$ and the same accretion 
rate ${\dot M}=0.1{\dot M}_{\rm Edd}$, where $M_B=10M_{\odot}$. The flow variables 
$log(v)$ (Figs. 7a, 7e, 7i), $log(T)$ (Figs. 7b, 7f, 7j), $\Gamma$ (Figs. 7c, 7g, 7k) 
and $log(n^{\prime}_e)$ (Figs. 7d, 7h, 7l) are plotted with $log(r)$. Here, 
$n^{\prime}_e=n_{e^+}+n_{e^-}$ is the total number density of leptons. In Figs. 7a, 7e, and 7i, the global 
solutions are presented by thick lines with arrow. The dotted lines in Figs. 7a and 7e, show ---
(i) the part of the
reflected-$\alpha$ topology, through which the global solutions which include shock waves do not pass, and
(ii) the part of the pre-shock branch which the flow would have continued if there was no shock. 
The dotted line in Fig. 7i, denote the wind type solution associated with the critical point. The dotted curves are subsequently dropped for variables other than $v$. The vertical jumps are the shocks.
Flows with $\xi \neq 0$ (Figs. 7a-7h), have two physical $r_c$, and undergo shock transitions 
at $x_s=423.46$ (for $\xi=1.0$) and $x_s=68.78$ (for $\xi=0.5$), respectively. 
As is evident from the Figures, even if both the flows containing protons admit shock waves
for the same flow parameters, the shock location and the strength will vary considerably. The \ep
flow is slower and hotter albeit less relativistic, and is of less $n^{\prime}_e$ than the $\xi=0.5$ flow.
Even though solved at the same ${\cal E}$ and $\lambda$, the \ee flow (Figs. 7i-7l) has only one critical point,
and therefore cannot have a shock transition.
Apart from the absence of shocks, the \ee flow is the slowest
(Fig. 7i), the coldest (Fig. 7j), thermally the least relativistic (Fig. 7k) and with the highest
$n^{\prime}_e$ (Fig. 7l).
Since the emitted radiation will depend on $n^{\prime}_e$, $T$, and $v$, apart from 
the gravitational redshift due to the central mass, Figs (7a-7l) suggest that the flows with different composition
even at the same ${\cal E}$, $\lambda$ and ${\dot M}$ will produce dramatically different spectrum.

\begin{figure}[h]
\begin{center}
{
\includegraphics[scale=0.7]{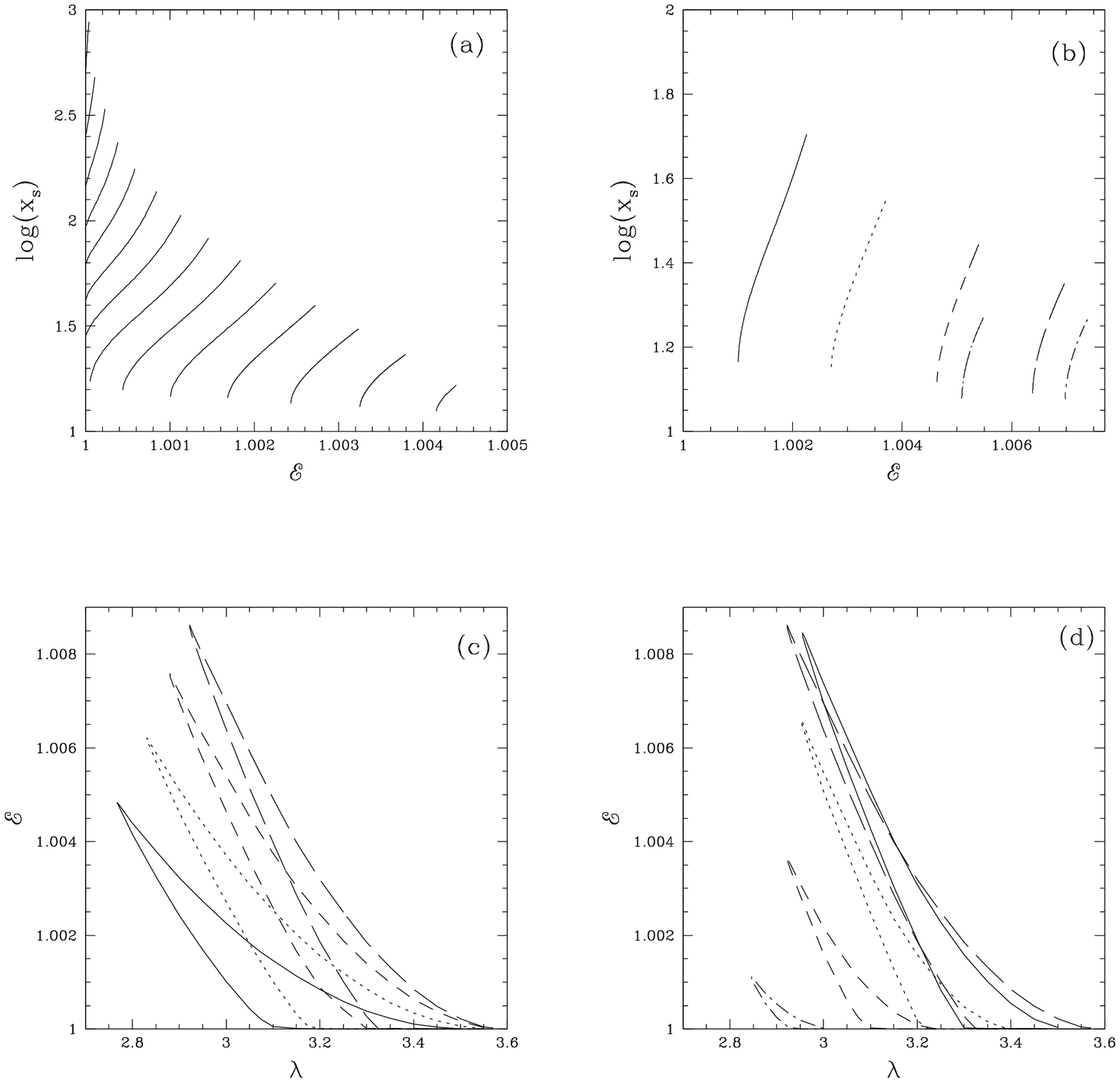}
}
\end{center}

\caption{(a) Variation of $log(x_s)$ with ${\cal E}$ for \ep flow. The leftmost curve
is for $\lambda=3.45$ and decreases by $d\lambda=0.05$
for each curve towards right, up to $\lambda=2.8$. (b) Variation of $log(x_s)$ with ${\cal E}$ for $\lambda=3$, and each curve is for $\xi=1.0$ (solid), $0.8$ (dotted), $0.6$ (dashed), $0.4$ (long dashed), $0.2$ (dashed),
$0.1$ (long dashed dotted). (c) The shock domain of  ${\cal E}-\lambda$ parameter space for
$\xi=1.0$ (solid), $0.8$ (dotted), $0.6$ (dashed), $0.4$ (long-dashed).
(d) The shock domain of  ${\cal E}-\lambda$ parameter space for
$\xi=0.2$ (solid), $0.1$ (dotted), $0.05$ (dashed), $0.025$ (dashed-dot).
$\xi=0.4$ (long dashed) has been plotted for the sake of comparison.}
\label{fig8}
\end{figure}

In Fig. 8a, variation of $log(x_s)$ with ${\cal E}$ is plotted for \ep flow.
The left most curve is for $\lambda=3.45$ and $\lambda$ decreases by $0.05$ for each curve towards the right.
At a given $\lambda$, $x_s$ increases
with ${\cal E}$, while at a given ${\cal E}$, $x_s$ decreases with the
decreasing $\lambda$. In other words, a flow with the lesser $\lambda$ 
needs more energetic flow to produce a strong enough barrier 
that can cause a shock transition. This is the general trend of a flow 
at a given value of $\xi$ ($>0$). In Fig. 8b, $x_s$ is plotted with 
${\cal E}$ for $\lambda=3$, and each curve denotes $\xi=1.0$ (solid),
$0.8$ (dotted), $0.6$ (dashed), $0.4$ (long dashed), $0.2$ (dashed dotted),
$0.1$ (long dashed dotted). For a given $\lambda$, the effect of rotation is the same for all types of flows. 
So the difference in the solution arises only due to the difference in the thermal 
state of the flow. Although the reduction of $\xi$ results in colder flow, 
it also reduces the rest energy per pair of oppositely charged particles. 
In the range $1.0 >\xi{\gsim} 0.24$, a decrease of $\xi$ results in a cooler 
but more relativistic and energetic flow, and hence the shocks
are formed closer to the black hole and for higher ${\cal E}$. For $0.24 > \xi \geq 0.0$, a reduction of $\xi$ reduces 
the thermal energy to the extent that the associated reduction in rest energy cannot compensate for the 
reduction of the former. This results in a flow which is less relativistic and less energetic.
Hence, the shock shifts to the region of lesser ${\cal E}$. 
Parameter space for a shock shows a similar behaviour. Shock domains in the ${\cal E}-\lambda$ space are plotted for
$\xi=1.0$ (solid), $0.8$ (dotted), $0.6$ (dashed), $0.4$ (long-dashed) in Fig. 8a, and those
for $\xi=0.4$ (long-dashed), $0.2$
(solid), $0.1$ (dotted), $0.05$ (dashed), and $0.025$ (long dashed) are plotted in Fig. 8b.
Hence it is clear that the shock domain shifts to higher ${\cal E}-\lambda$ region, 
as $\xi$ is reduced in the range $1\geq \xi{\gsim} 0.24$, but then to lower ${\cal E}
-\lambda$ region as $\xi$ is reduced in the range $0.24 > \xi > 0$. The region under the 
curve also shrinks with the decrease of $\xi$ in Fig. 8d, and finally vanishes for $\xi=0.0$.

\section{Discussion and concluding remarks}

Accretion solutions onto black holes are necessarily transonic and close to the 
horizon they are definitely sub-Keplerian. It has been shown that the thermodynamics 
of relativistic flow depends on its composition if the temperature range is $10^8K<T<10^{13}K$ (\ie CR09).  
Since the temperature of rotating-relativistic transonic flows happen to be in the same range, 
the relativistic thermodynamics of CR09 has been extended to such flows in this paper.

It has been shown in this paper that the multiplicity of physical critical points is not realized 
for \ee flow, for any value of $\lambda$ (\eg Figs. 3c, 4). Multiplicity of physical 
critical points is achieved only for flow with $0< \xi \leq 1$. The domain for multiple 
critical points in ${\cal E}-\lambda$ space (\ie MCP region) 
moves to the right \ie to higher ${\cal E}$ and $\lambda$ with the decrease of $\xi$ 
in the range $1 \geq \xi \gsim 0.24$. Further decrease of $\xi$, in the 
range $0.24 \gsim \xi > 0.0$ causes the MCP region to move to the 
left \ie to lower ${\cal E}$ and $\lambda$. In this range, the region
under the curve MCP shrinks with $\xi$ and ultimately 
becomes zero for $\xi=0$. Only $r_{ci}$ exists for \ee, which means the solution 
will be subsonic throughout and become supersonic only very close to the horizon, 
\ie for \ee flow the solution is I type, independent of the value of $\lambda$.
Different types of accretion solutions (\eg I, W, A, or O), can be 
achieved only for flows with $1\geq \xi > 0$ (see, Figs. 6a-6c).

One of the consequences of A type solutions is the formation of accretion shocks. 
Shocks in accretion have been invoked to explain the spectrum (Chakrabarti \& Titarchuk, 1995;
Chakrabarti \& Mandal, 2006; Mandal \& Chakrabarti, 2008),
formation of jets and outflows (Das \& Chakrabarti, 1999; Fukumura \& Kazanas, 2007; 
Das, Becker \& Le, 2009) etc. We show that
along with the conserved quantities such as ${\cal E}$ and $\lambda$, and the dissipative effects,
the composition of the flow is also important to decide 
whether a shock would eventually form or not. A flow composed of 
similar particles like \ee flow, is unlikely to form a accretion shock around black holes. 
Even for flows with $1\geq\xi>0$, at the same \{${\cal E},~\lambda$\}, 
the global solution as well as the shock location will be different for 
different $\xi$ (Figs. 7a-7l, 8b-8d).

As matter falls into the BH, the gravitational energy released will enhance both the 
kinetic and the thermal energy of the flow. The models which concentrate only on the regime 
where the gravitational energy is converted mainly to the thermal energy, may either be 
cooling dominated (\eg SS disk; Shakura \& Sunyaev, 1973; Novikov \& Thorne, 1973) or advection dominated (Narayan, Kato \& Honma, 1997).
Either way, these models remain mainly subsonic (except very close to the black 
hole) and hence do not show shock transition. If one scans the entire
parameter space then one retrieves solutions which have 
a kinetic energy so much so that it becomes transonic at distance of few $\times ~ 100r_g$.
A subset amongst these solutions admit shock transitions when shock conditions
are satisfied. This is the physical reason why
some disk models show shock transitions and others do not (Lu, Gu \& Yuan, 1999).

In this paper, we ignored the explicit forms of dissipation, heating, cooling,
and radiative transfer.  All of these are likely to modify the parameter space in which 
the shock solutions were found. In several papers we have already discussed the 
direction in which the admissible parameter space is likely to be shifted (e.g., Chakrabarti \& Das, 2004).
Presently, we find that in the range, $1 \geq \xi \gsim 0.24$, the parameter space moves to the right 
\ie higher ${\cal E},~\lambda$, while in the range $0.24 \gsim \xi > 0$ the parameter space moves 
to the left \ie lower ${\cal E},~\lambda$ (\eg Figs. 8c-8d).  However, with the increase
of the viscosity parameter the shock parameter space moves to the left (Chakrabarti \& Das, 2004; Gu \& Lu, 2004; Das, 
Becker \& Le, 2009). This implies that the combination of composition and viscosity 
might result in shock parameter space to be located at higher angular momentum, than, say, purely \ep flow. 

Along with the heating and the cooling processes, one needs to be concerned about the spectral and timing
properties from these flows. The spectrum will clearly depend on $\xi$. 
The spectrum for \ee flow at the same \{${\cal E}, \lambda$\} will be
significantly more different compared to the solutions of flow with $\xi \neq 0$. 
The \ee flows will form in presence of very strong pair-production. 
It will have a very high opacity and would cool faster in presence of 
seed photons. In presence of inverse-Comptonization, some regions of the \ee
flow may rapidly cool and force the flow to become supersonic and will not have shocks.
However, observations of line emissions in 
jets and outflows have indicated the presence of baryons (e.g. SS433) and it
is likely that the accretion (coming from a companion star) and the winds 
will have significant baryons. Even if \ee outflows form, baryon loading 
on the way will enhance proton fractions and eventually shocks may also form.
In other words, even the composition may and should vary on the way. This requires
a full treatment of the problem which is beyond the scope of this paper and will be dealt with
separately.

\end{document}